\+

# Nitrogen flow rate dependent atomic coordination, phonon vibration and surface analysis of DC Magnetron sputtered Nitrogen rich-AlN thin films

Aishwarya Madhuri[a], Sanketa Jena[a], Mukul Gupta[b], Bibhu Prasad Swain[a,*]

[a]National Institute of Technology Manipur, Langol Rd, Lamphelpat, Imphal, Manipur 795 004, India

[b]UGC-DAE Consortium for Scientific Research, University Campus, Khandwa Road, Indore, 452 001, India

*Corresponding author: bibhuprasad.swain@gmail.com

**ABSTRACT**

In this work, the effect on crystallite orientation, surface morphology, fractal geometry, structural coordination and electronic environment of DC magnetron sputtered AlN films were investigated. X-ray diffraction results disclosed that the c-axis orientation of AlN films increased with the preferred wurtzite hexagonal structure above 17% $N_2$ flow. X-ray reflectivity data confirmed AlN film density increased with increasing $N_2$ flow and was found to be 3.18g/cm$^3$ for 40% $N_2$. The transition of electrons from N *1s* to *2p* states hybridized with Al *3p* states because of $\pi^*$ resonance was obtained from X-ray absorption spectroscopy of the N K-edge. The semi-empirical coordination geometry of nitrogen atoms has been studied by deconvolution of N K-edge. The surface composition of AlN films at 40% $N_2$ consists of 32.08, 51.94 and 15.97at.% Al, N and O respectively. Blue-shifting of $A_1$(LO) and $E_1$(LO) modes in the Raman spectra at phonon energies 800 and 1051cm$^{-1}$ respectively was most likely due to the presence of oxygen bonds in the AlN films.

+

## 1 Introduction

Among the nitride family, AlN has a broad bandgap of 6.12 eV (~200 nm) at room temperature. Hexagonal AlN has high acoustic velocity (11 km/s for longitudinal waves) [1], a hardness of more than 20 GPa and a melting temperature of 2200°C. AlN is used as a dielectric layer and a heat spreading layer due to its excellent electrical resistivity of $10^{14}$ Ω with a dielectric constant (k=8.5) and a thermal conductivity of 321 W $m^{-1}$ $K^{-1}$ [2], respectively. Because of high optical transparency, electrical and thermal properties, AlN thin films have many applications in numerous fields such as microelectromechanical system sensors (MEMS) [3], bulk acoustic resonators (BAR) [4], energy harvesters [5], scanning micro-mirrors [6], deep ultra-violet devices [7], microphones [8] and pressure sensors [9]. AlN is an appropriate piezoelectric transducer material with a piezoelectric co-efficient of 6 $pC.N^{-1}$ [10,11]. The crystal orientation and piezoelectric coefficient are directly linked with each other. AlN has no hysteresis, curie temperature and aging effects as it is ferroelectric [12].

To yield high-quality thin films, ease of scale up and increase in the deposition rate, magnetron sputtering is widely used for the last few decades for thin film deposition. Cho et al. [13] reported that at 10% $N_2$ flow, the AlN thin films show a strong c-axis orientation. Signore et al. [14] found out that the crystallinity and orientation of AlN thin films are along the c-axis perpendicular to the plane of the substrate by increasing the $N_2$ flux percentage in rf magnetron sputtering. Ababneh et al. [15] reported a decrease in the intensity of the (002) peak of AlN thin films deposited using DC MS with the increase in working pressure during sputtering and widening of the corresponding full width at half maxima. Venkataraj et al. [16] found that the films prepared at low $N_2$ flow (less than 2 sccm) were crystalline and in between 2 to 5 sccm $N_2$ flow films were amorphous and above 5 sccm hexagonal wurtzite structure of the films were formed.

Though the above researchers investigated the structural and optical properties of AlN thin films deposited by RF magnetron sputtering, however, the phonon vibration, bonding coordination, surface roughness and density of AlN thin films using DC reactive magnetron sputtering for the deposition have not been studied properly. Therefore, the present work is aimed at the followings: (a) the role of nitrogen flow in structural properties, (b) the investigation of the morphology and fractal geometry, (c) the investigation of phonon vibration and structural co-ordination, (d) moreover, it also describes the local geometry and atomic coordination of AlN with $N_2$ flow rate, and (e) formation of oxide layer on the surface of AlN films and chemical bonding behaviour of Al and N were also explained. In this paper, the AlN thin films were deposited by the dc magnetron sputtering technique on Si(100), quartz and glass substrates.

## 2  Experimental details

### 2.1  Film depositions

The AlN thin films were deposited on single-side polished p-type Si (100), quartz and glass (2 cm × 2 cm) substrates by using a dc magnetron sputtering system (AJA Int. Inc.). A 3-inch diameter Al target with a purity of 99.999% was kept on a water-cooled magnetron cathode. The silicon and quartz substrates were cleaned in an ultrasonic bath of acetone for 10 minutes and then with isopropanol, while the glass substrates were cleaned in water (soap solution) first and then with isopropanol. The distance between the substrates and the target was kept at 120 mm. Evacuation of the

+

main chamber to a base pressure of $4\times10^{-7}$ Torr or lower using a TMP backed by a rotary pump was done before the Ar and $N_2$ gas (both gases were 99.999% pure) were introduced and the working pressure varied accordingly (~$10^{-3}$ Torr). The total gas flow (Ar+$N_2$) was maintained at 50 sccm while the Ar and $N_2$ gas flow percentage were varied being controlled by the mass flow controller. The deposition was carried out at a fixed DC power of 200 W by rotating the substrate holder at 60 rpm for uniform growth of the film during the whole deposition process. No additional heating was applied to the substrates during the film deposition. During the deposition process, the colour of the AlN films went from transparent to metallic silver with the increase in the $N_2$ flow indicating the insulating behaviour changed to conducting behaviour of the AlN films.

**2.2 Film characterizations**

The lattice spacing (d), the crystallite size (D) and the micro-strain (ε) of the AlN thin films were estimated by using X-ray diffraction (Bruker D8 Advance- Powder XRD) with Cu-$K_\alpha$ radiation (λ= 1.54 Å). An atomic force microscope (Bruker-made Bio-AFM) was employed to measure the RMS surface morphologies (Rrms) and fractal analysis of the AlN thin films surface where the AFM images were recorded in a non-contact tapping mode. Using X-ray reflectivity (Bruker D8 Discover), the rate of the film growth and mass density were calculated. Soft X-ray Absorption Spectroscopy at a synchrotron radiation source (BL-01, Indus-2, RRCAT) [17] was used to measure the N-absorption edges of AlN thin films. The beam energy of the synchrotron source was set at 2.5 GeV and the beam current was 120 mA. The AlN thin films were attached to the photocathode of the electron multiplier using adhesive carbon tape and contact between film and photocathode was made by using silver paste. The N K-edge spectra were observed in a total electron yield (TEY) mode under UHV conditions using a

+

simple drain-current setup. Under these conditions, the beamline energy resolution was about 0.2 eV. X-ray photoelectron spectroscopy was carried out using ESCA spectrometer (SPECS, Germany make) to calculate the surface composition of AlN films using nonmonochromatic Al K$_\alpha$ X-rays (hυ=1486.61 eV) and chemical bonding behaviour of Al and N were obtained from high-resolution XPS spectra of Al(2p) and N(1s) core orbital spectra. C *1s* peak (284.6 eV) was used as the reference peak for the correction of the charging effect in the XPS spectra. The AlN film surface was etched for 30 minutes using Ar ions with energy 3 KeV and ion current 0.5 µA to remove surface contaminants allowing for more accurate surface chemistry of AlN films. Raman spectroscopy studies were performed using Horiba Jobin Yvon (HR800) Raman spectrometer to measure the phonon vibration modes of AlN thin films. The excitation energy of the laser source used was 472 nm diode laser and the resolution of the spectroscopy was 1 cm$^{-1}$.

## 3  Results and discussion

### 3.1  Atomic force microscopy

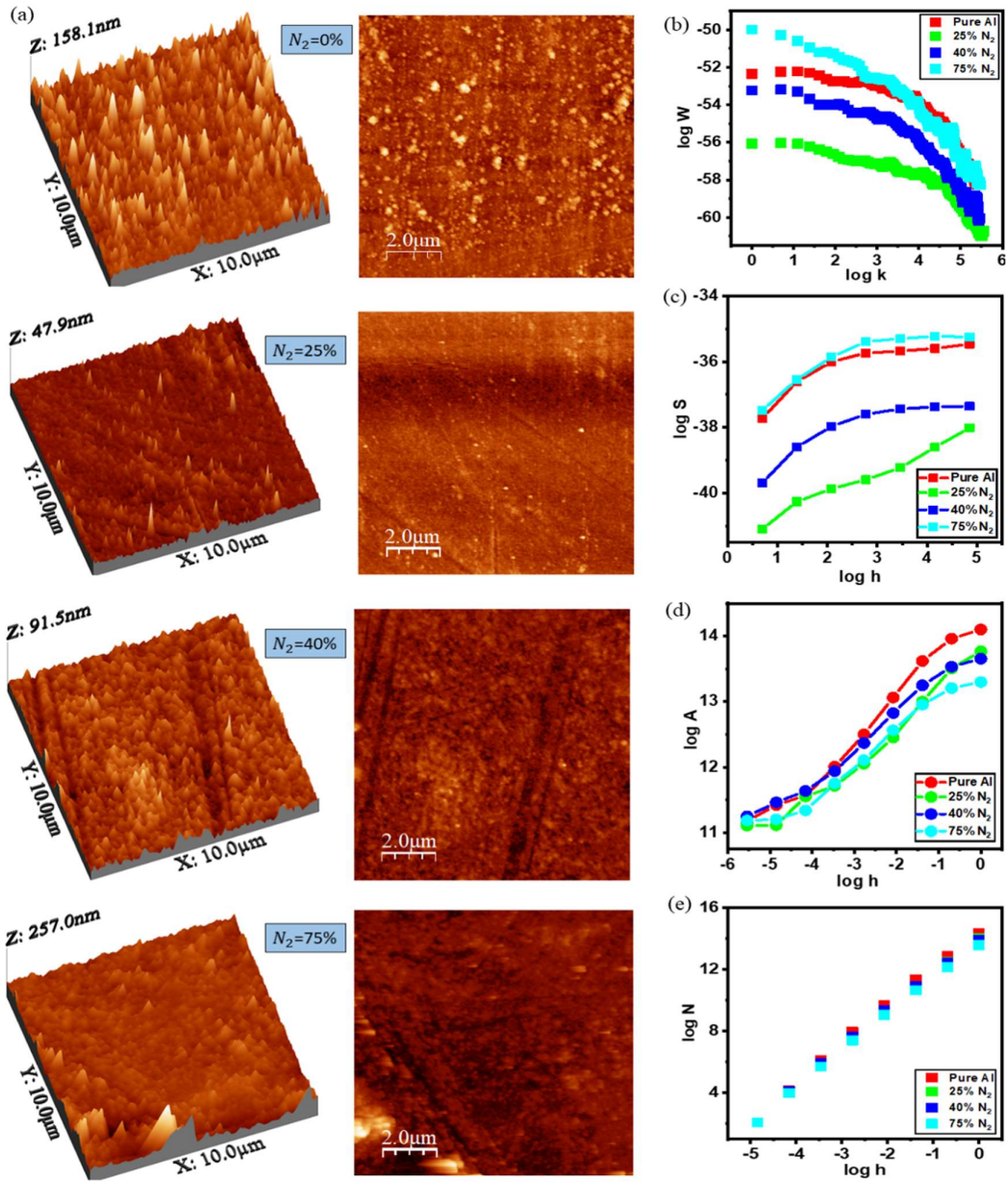

[**Fig. 1(a)** AFM images of AlN thin films with $N_2$ flow rate 0%, 25%, 40% and 75% and fractal analysis by **(b)** power spectral density **(c)** Partitioning **(d)** Triangulation **(e)** Cube counting.]

Fig. 1 shows the AFM images (10×10 μm$^2$) of the AlN films deposited on glass substrates at 0, 25, 40 and 75% $N_2$ flow rates. AFM images give information about the surface roughness and surface characteristics such as the fractal geometry of thin films.

+

The contrast of the AFM images varies from 47.9 nm to 257 nm with the variation in N$_2$ %. The roughness average (Ra) and root mean square surface roughness (RMS) have been calculated using WSxM 5.0 software shown in Table 1 below. For 75% N$_2$, RMS roughness is the highest at 16.336 nm for 75% N$_2$ flow and pure Al thin film at 16.045nm.

| N$_2$ (%) | RMS roughness (nm) | Roughness average (nm) |
| --- | --- | --- |
| 0 | 16.04±0.64 | 11±0.44 |
| 25 | 2.89±0.14 | 2±0.10 |
| 40 | 5.71±0.22 | 4±0.20 |
| 75 | 16.33±0.65 | 10±0.40 |

[**Table 1** Calculated RMS roughness and roughness average for AlN thin films with various N$_2$ flows including the error bars.]

The fractal dimensions of AlN thin films were evaluated by Gwyddion software with triangulation and cube-counting; based on box-counting fractal dimension and power spectral density and partitioning; based on power spectrum dependence of fractional Brownian motion as shown in Fig.1.

The optical scattering concentrates on the calculation of the bidirectional reflectance distribution function (BRDF) which allows to determine power spectral density (PSD) function for the characterization of topography surfaces and films. Power spectral density is described in the spatial frequency domain which is defined as the roughness power per unit frequency over the length of the sample. PSDF is calculated from the surface profile h(r) being Fourier transformed [18]. PSDF for 1-D surface profile is defined as;

$$\text{PSD}(f_r) = \lim_{L \to \infty} \langle \left| \frac{2}{L} \int_{-L/2}^{L/2} h(r) e^{i2\pi f} \, dr \right|^2 \rangle \quad (1)$$

Where the spatial frequency and the sampling length are denoted by f and L respectively.

From the PSD function, the roughness can be directly calculated by using the integral;

$$\sigma^2 = \int_{f_{min}}^{f_{max}} PSD(f) \, df \qquad (2)$$

For thin films, the bidirectional reflectance distribution function (BRDF) depends on the topographies of the upper and bottom surfaces of the thin films. If the upper and lower surfaces of the films are identical then the PSD function is given by;

$$BRDF = |F_1 + F_2|^2 \, PSD \qquad (3)$$

Where $F_1$ and $F_2$ are the optical functions of air-film and film-substrate interfaces respectively.

In the case of non-identical surfaces, the BRDF is expressed by

$$BRDF = |F_1|^2 \, PSD_1 + |F_2|^2 \, PSD_2 \qquad (4)$$

Where $PSD_1$ and $PSD_2$ correspond to the power spectral densities of upper and lower film surfaces respectively.

The fractal dimension in fig-2b representing power spectral density varied from 2.669 to 2.526 with the increase in $N_2$ flow. Fig-2c shows the log-log plot of the partition function which varies from 2.765 to 2.746. Fig-2d, e shows the triangular and cube box-counting of AlN thin films at various $N_2$ flows. The details of the fractal dimension are shown in the table-2 below.

| $N_2$ flow % | Power spectrum | Partitioning | Triangulation | Cube-counting |
|---|---|---|---|---|
| 0 | 2.669 | 2.765 | 2.589 | 2.526 |
| 25 | 2.889 | 2.660 | 2.516 | 2.457 |
| 40 | 2.634 | 2.743 | 2.480 | 2.429 |
| 75 | 2.526 | 2.746 | 2.445 | 2.369 |

+

[**Table 2** Calculated fractal factors from the Power spectrum, Partitioning, Triangulation and Cube-counting.]

### 3.2 X-ray diffraction

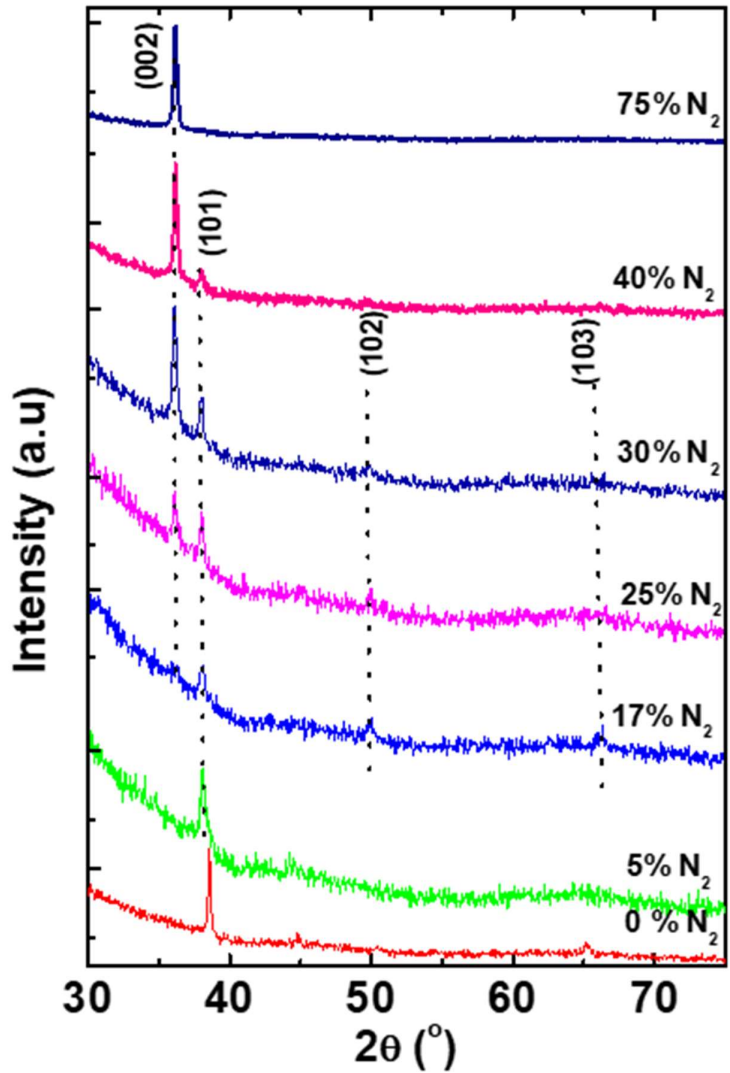

[**Fig. 2a.** XRD patterns of AlN thin films deposited on quartz substrates by varying $N_2$ flow percentage.]

The X-ray diffraction (XRD) spectra of AlN thin films with different $N_2$ flow percentages keeping the total flow rate of 50 sccm is shown in Fig. 2a. The thickness of the AlN thin films was maintained at 200 nm. For pure Al thin films (metal-rich growth mode), the diffraction peaks are observed at 38.5º, 44.7º and 65.1º corresponding to (111), (200), and (220) planes respectively. But then with the introduction of reactive gas ($N_2$), the metal-rich mode changed to reactive gas-rich mode and the deposition rate decreased exponentially as shown in Fig. 2b which was also observed by Tatejima et al. [19].

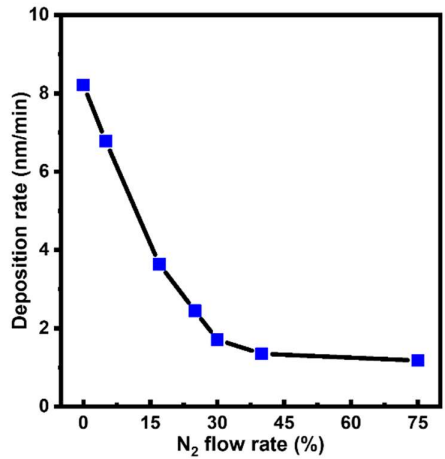

[**Fig. 2b.** Graph between deposition rate of AlN thin films with respect to $N_2$ flow percentage.]

From 5 to 30 % $N_2$ flow rate, four diffraction planes were observed at 36º, 38.08º, 50.04º and 66.3º corresponding to (002), (101), (102) and (103) planes respectively indicating the polycrystalline nature of AlN thin films. Okano et al. [20] and Liu et al. [21] showed that with the decrease in $N_2$ flow rate, the (002) crystallographic orientation increases and only the (002) plane appeared for 40% $N_2$ flow. At 40% $N_2$ concentration and 120 mm separation between the Al target and substrate, Ishihara et al. [22] observed a strong (100) plane peak and a weak (101) peak. However, with an increase of $N_2$ flow percentage above 40%, (101), (102) and (103) planes are not preferred planes for AlN growth, whereas (002) is the preferred plane. The peak intensity of (002) gradually increases indicating the growth of the film oriented along the c-axis which confirms the Wurtzite-type structure of AlN with the increase in $N_2$ flow percentage. This may be due to the target sputtering mode changing from transition mode to the reactive mode during the deposition with the increase in $N_2$ flow rate. The crystallite size of AlN thin films has been calculated using the Debye-Scherrer equation;

$$D = \frac{k\lambda}{\beta \cos\theta} \tag{5}$$

Where D is the crystallite size (in nm), k is the shape factor varied from 0.65 to 2, $\lambda$ is the wavelength of X-ray (0.154 nm), $\beta$ is the full width at half maxima in radian and $\theta$ is the angle of diffraction.

The dislocation density is evaluated using Williamson–Smallman equation;

$$\delta = \frac{1}{D^2} \tag{6}$$

Where $\delta$ is the dislocation density.

Using XRD results the strain is calculated by the given relation;

$$\varepsilon = \frac{\beta \cot\theta}{4} \tag{7}$$

For the pure Al thin film, the crystallite size is about 40 nm while decreasing to about 30 nm with an increasing $N_2$ flow rate. The dislocation density varied from $6.04 \times 10^{14}$ m$^{-2}$ to $1.18 \times 10^{15}$ m$^{-2}$ with increasing $N_2$ fraction. Fig. 2c shows the variation in crystallite size and micro-strain with the $N_2$ flow rate. Table 3 shows the calculated lattice parameters and dislocation densities for AlN thin films. The lattice parameters could not be calculated for AlN thin films with 5% and 17% $N_2$ flows as the AlN thin films growth was oriented in both the *a* and *c*-axis for these $N_2$ flow.

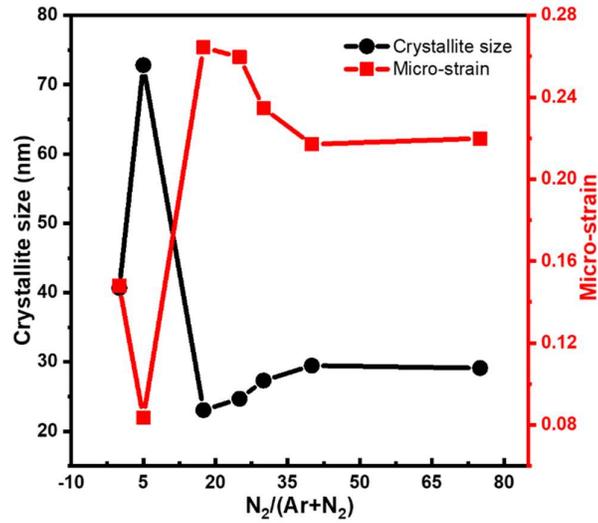

[**Fig. 2c.** Graph between crystallite size and micro-strain vs N$_2$ flow rate.]

| | | **This work** | | | **Literature** | |
|---|---|---|---|---|---|---|
| RN$_2$ (%) | (hkl) | d-spacing (A°) | Lattice parameters (A°) | Dislocation density (1/m$^2$) | Lattice parameters (A°) | Reference |
| 0 | (111) | 2.33±0.11 | a = 4.04±0.13 | 6.04×10$^{14}$ | c=5.0207 | Ababneh et al. [15] |
| 5 | (101) | 2.36±0.11 | --- | 1.88×10$^{14}$ | c = 4.98 | Okano et al. [20] |
| 17 | (101) (102) (103) | 2.35±0.09 1.82±0.07 1.40±0.05 | --- | 1.88×10$^{15}$ | c= 4.98 | Liu et al. [21] |
| 25 | (101) (002) | 2.36±0.11 2.48±0.09 | a=3.10±0.11 c=4.97±0.17 | 1.64×10$^{15}$ | a = 3.109 c = 5.021 | Alsaad et al. [23] |
| 30 | (002) (101) | 2.48±0.12 2.36±0.09 | c=4.97±0.19 a=3.09±0.12 | 1.34×10$^{15}$ | c= (4.978 ± 0.001) a= (3.113 ± 0.001) | Gablech et al. [24] |
| 40 | (002) (101) | 2.48±0.09 2.36±0.11 | c=4.96±0.16 a=3.10±0.11 | 1.15×10$^{15}$ | c = 4.98 | Aissa et al. [25] |
| 75 | (002) | 2.36±0.11 | c=4.96±0.17 | 1.18×10$^{15}$ | c = 4.980 a = 3.110 | Schulz et al. [26] |

[**Table 3** Calculated lattice parameter and dislocation density including the error bars from XRD data.]

+

## 3.3 X-ray reflectivity

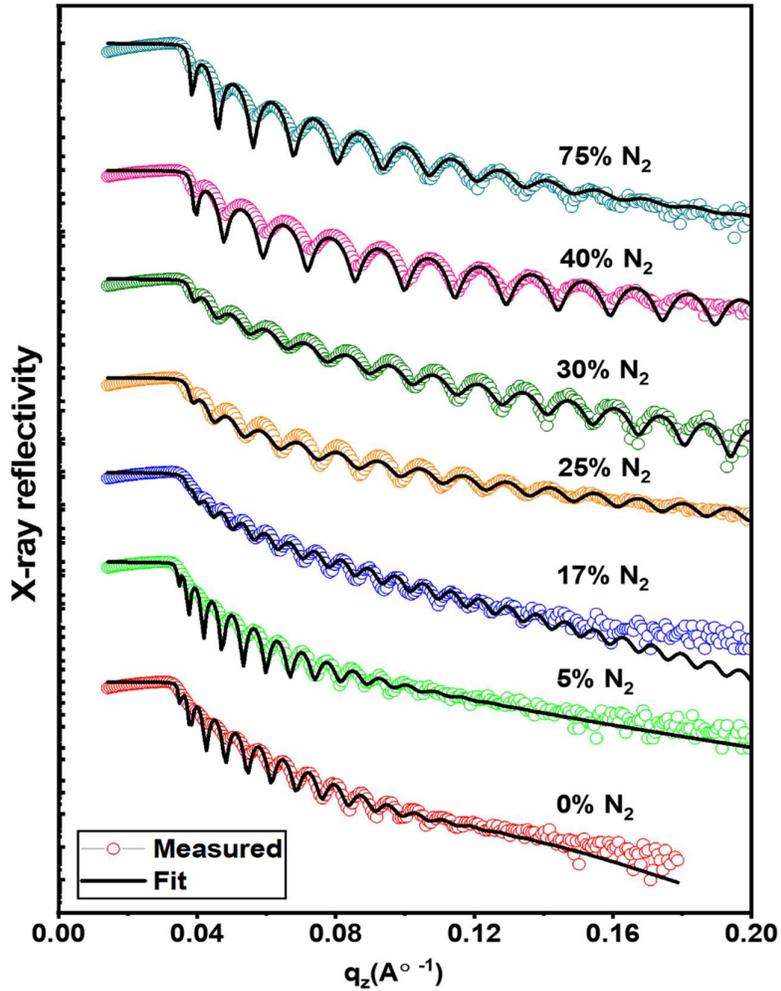

[**Fig. 3.** Measured (circles) and fitted (solid line) XRR patterns of optimized AlN thin films on Si substrates with the variation in $N_2$ flow percentage.]

Fig. 3 shows the X-ray reflectivity (XRR) plot with different $N_2$ fractions of the optimized AlN thin films. The measurements were performed using a Bruker D8 to discover a system with Cu $k_\alpha$ rays ($\lambda$= 1.54 Å) on AlN thin films deposited on Si substrates.

For total reflection the critical angle $\Theta_C$ is given by;

+

$$\Theta_C = \sqrt{2\delta} \tag{8}$$

Where δ is the dispersion given by;

$$\delta = \frac{r_e \lambda^2}{2\pi} N_o \rho \frac{\sum_i x_i(z_i + f'_i)}{\sum_i x_i M_i} \tag{9}$$

where $r_e$ is the classical radius of an electron (2.818×10⁻⁹m), $N_o$ is the Avogadro no. (6.022×10²³ mol⁻¹), ρ is the density (g/cm³) and for $i^{th}$ atom; $z_i$ is the atomic no., $M_i$ is the atomic weight, $x_i$ is the atomic ratio, $f'_i$ is the atomic scattering factor.

So, the AlN thin film density is estimated from the above relation.

The distance between the adjacent interference maxima is approximated by:

$$\delta\alpha_i \approx \frac{\lambda}{2t} \tag{10}$$

where it is the film thickness.

From the slope of the reflectivity, the film roughness was calculated.

The critical angle of the films varies from 0.45° to 0.48° with the increase in the $N_2$ fraction indicating the density of the AlN thin films increases with an increase in the $N_2$ fraction. For pure Al, the calculated density of 2.78 g/cm³ well matches the actual value which is 2.7 g/cm³. However, when the $N_2$ flow rate was changed, the density of the AlN thin films also changed accordingly, and at 40% $N_2$ flow rate, the density was up to 3.18 g/cm³, which is very close to the actual AlN density 3.26 g/cm³. When the nitrogen flow rate is increased, there is an increased likelihood of ion-ion collisions that led to ion-energy exchange. Consequently, the average energy of the ions in the plasma increased. When these energetic ions bombarded the substrate, they delivered more energy and momentum to the substrate surface which can contribute to densification and reduction in roughness of the AlN films. The deposition rate of AlN thin films

+

decreases with the increase in N₂ flow. Also, the roughness of the films was found to be the smallest for 40% N₂ flow and highest for pure Al thin films as shown in table 4 below.

| RN$_2$ (%) | Thickness (nm) | Density (g/cm$^3$) | Roughness (A$^o$) |
|---|---|---|---|
| 0 | 73.94±1 | 2.78±0.08 | 13.78±1 |
| 5 | 74.61±2 | 2.69±0.10 | 6.52±2 |
| 17 | 73.12±1 | 3.15±0.09 | 11.08±1 |
| 25 | 47.96±1 | 3.03±0.12 | 9.24±2 |
| 30 | 46.19±2 | 3.17±0.09 | 7.69±2 |
| 40 | 40.53±2 | 3.18±0.13 | 2.76±1 |
| 75 | 43.61±1 | 3.01±0.11 | 5.0±2 |

[**Table 4** Calculated thickness, density and roughness of AlN thin films from XRR including the error bars.]

### 3.4 XAS Analysis

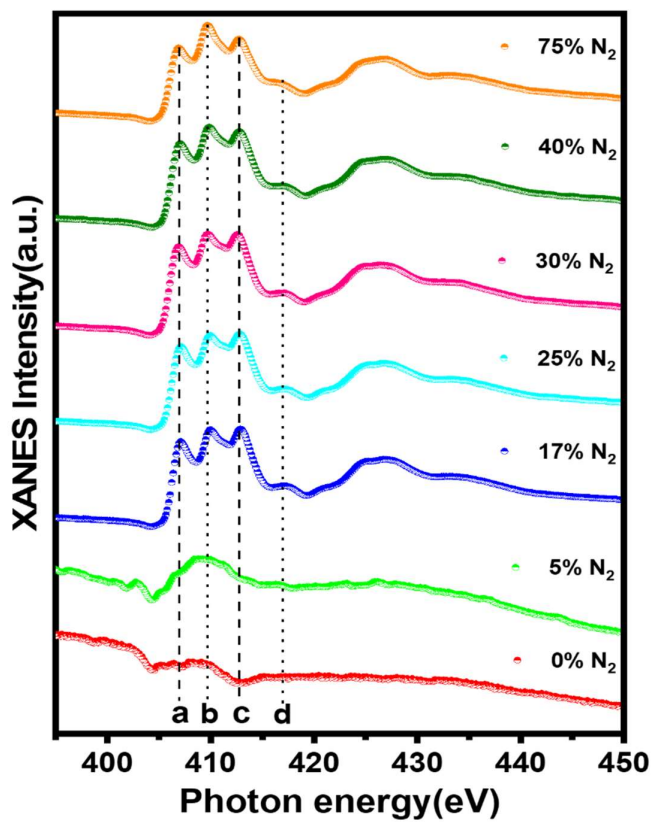

+

[**Fig. 4a.** SXAS spectra of N K-edge of AlN samples deposited at various $N_2$ flow rates.]

Fig. 4a shows the XAS of nitrogen K-edge of AlN thin films. XANES is the reflection of the density of unoccupied states of the absorbing atoms and the spectra contain information about local geometry and oxidation state. The nitrogen K-edge spectra for the transition of photon from *1s* to *2p* orbital were collected in a total electron yield (TEY) mode. After normalizing the XAS data the edge energy appeared to be at about 405 eV for AlN thin films which is 5 eV less than the theoretical value. The nitrogen K-edge spectra can be categorized in four different features such as *a*, *b*, *c* and *d* appearing at different photon energies for different $N_2$ flows. For pure Al, only one hump is observed indicating there is a small fraction of nitrogen present in the thin film with a fraction of impurity. For 5% $N_2$ flow features - *a*, *b* and *c* have been merged into one large peak showing the amorphous nature of AlN which is contradictory as it is crystalline confirmed from XRD data. After that up to 40% $N_2$ flow all the features that appeared are showing similar patterns. But for 75% $N_2$ flow intensity of feature-*b* is more than that of *c* which may be due to the only c-axis orientation of this AlN film. A large hump is observed at 426.38 eV from 17% to 75% $N_2$ flow which is absent in pure Al and 5% $N_2$ flow specifying the oxygen content is very low in both films.

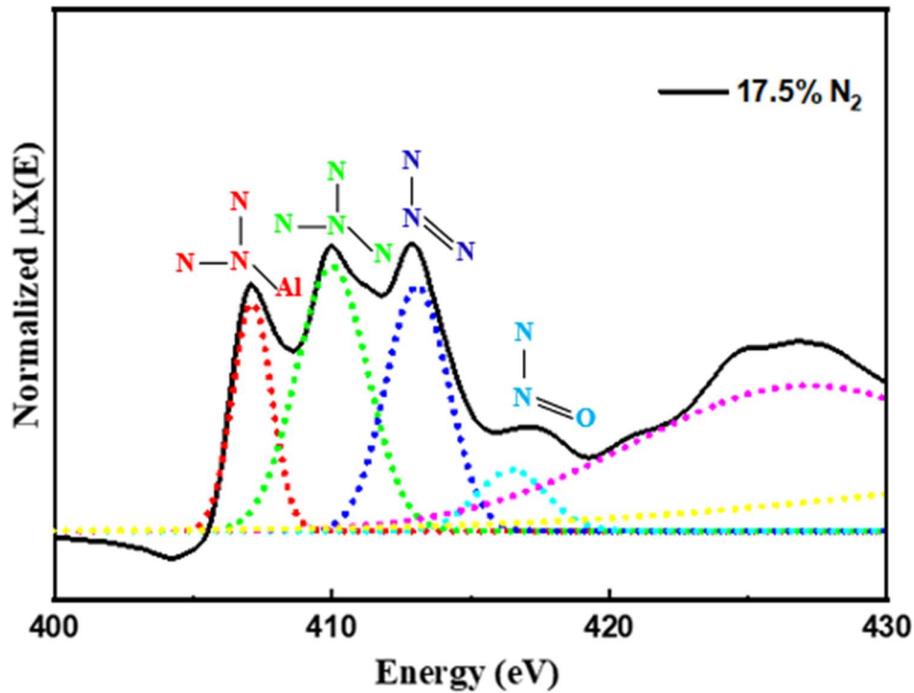

[**Fig. 4b.** Deconvoluted XAS spectra of N K-edge of AlN thin films at 17% $N_2$ flow showing the bonding coordination.]

Fig. 4b shows the deconvoluted XAS spectra of N K-edge by the Gaussian fitting method for 17% $N_2$ flow that represents the expected contributions from different coordination geometries in AlN films. This information provides insights into the local atomic structure around the N-atom. The above features correlate to the unoccupied nitrogen *2p* states that are hybridized with neighbouring Al *3p* states [27]. These features are assigned to various bonds due to the transition of electrons from 1s to *$2p_{x,y}$ ($\sigma^*$)* and *$2p_z$ ($\pi^*$)* orbitals. Features *a* and *b* are assigned to the transition of electron from N *1s* to the lowest unoccupied antibonding $2\pi_g$ level of N *2p* states hybridized with Al *3p* states due to $\pi^*$ resonance and features *c* and *d* are ascribed to the unoccupied N *2p* states hybridized with Al *3d* states for $\sigma^*$ resonance. An approach to all the possible semi-empirical coordination geometry of N atom with other atoms present in the AlN films was made. Features - *a*, *b*, *c* and *d* for 17% $N_2$ flow appearing at 407.58 eV, 410.64 eV, 413.01eV and 417.15 eV are assigned to N-Al, N-N, N=N and N-O states respectively [28]. However, the N-Al, N-N, N=N and N-O states correspond to the N atom bonded to Al, N and O atoms in the AlN films resulting in a decrease in electron density around the N-atom corresponding to electronegativity of the bonded element. Feature-*a* shifts from photon energy 407.58 eV to 406.96 eV with the increase in $N_2$ flow from 17% to 75% which may be due to the voids present in the AlN thin films. The absence of pre-edge in XAS spectra except for 5% $N_2$ flow indicates that there is no empty bound state near the Fermi level and fewer defects in the films. N K-edges did not appear for 0% $N_2$ flow and they started to appear with the increase in $N_2$ flow. Features *a*, *c* and *d* are prominent for 17% $N_2$ flow which can depict the fact that nitrogen started to accumulate during the deposition.

### 3.5 XPS analysis

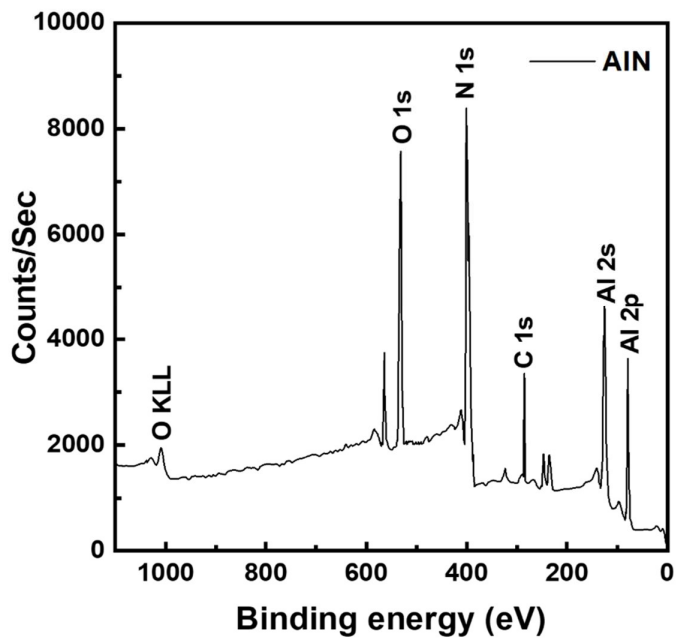

[**Fig. 5a.** Broad scan XPS spectrum of AlN thin films deposited at 40% $N_2$ flow.]

Fig. 5a shows the full scan XPS spectrum of AlN thin films deposited on Si substrate at room temperature for 40% $N_2$ flow. Five core orbital peaks were observed at 78.91, 124.85, 284.08, 400.48 and 531.82 eV binding energy corresponding to Al(2p), Al(2s), C(1s), N(1s) and O(1s) core orbitals respectively. Another peak corresponding to the Auger peak represented as O KLL appeared at 1007.15 eV. The reason behind the appearance of the C(1s) peak may be due to the manhandling impurity on the surface of the AlN films. The surface chemical composition for each element present in the XPS spectrum was calculated by using the following equation;

$$X_i = \frac{\left(\frac{A_i}{S_i}\right)}{\Sigma_i \left(\frac{A_i}{S_i}\right)} \qquad [11]$$

where $X_i$ is the atomic concentration of the $i^{th}$ element, $A_i$ is the integrated area and $S_i$ is the relative sensitivity factor of $i^{th}$ element on the surface of AlN films. The relative sensitivity factors used for the calculation of atomic percentages of Al, N and O are 0.537, 1.8 and 2.93 respectively. The calculated atomic concentrations of Al, N and O on the surface of AlN films are 32.08, 51.94 and 15.97 at.% respectively. Similarly, without taking O into account, the film composition can be calculated only by taking Al and N into consideration. The atomic percentages of Al and N within the AlN films are 38.18 and 61.81 at.% respectively.

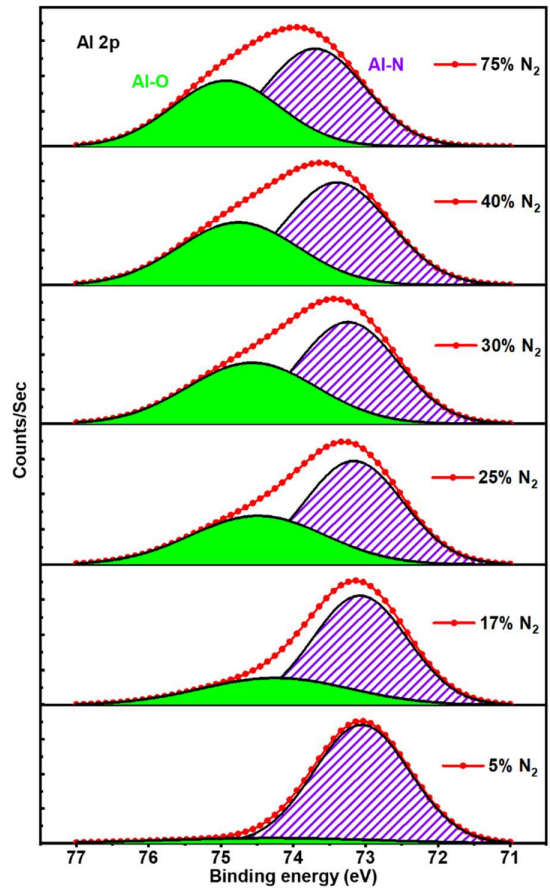

[**Fig. 5b.** Deconvolution of high-resolution XPS spectra of Al(2p) core orbital of AlN thin films at a different $N_2$ flow rate.]

The deconvolution of Al(2p) and N(1s) core orbital implies valuable significances such as the surface signature of AlN films and the chemical bonding behaviour of Al with Al and N and N with Al, N and O atoms. The indexing of the deconvoluted peaks was done by the difference in electronegativity and the possibility of chemical bonding with the constituent elements. Fig. 5b shows the deconvolution of Al(2p) core orbital of AlN films at different $N_2$ flow percentages. The deconvoluted Gaussian-Lorentzian subpeaks were assigned to Al-N (nitride) and Al-O (oxide) bonds and peaked at 73.2 and 74.4 eV respectively which are in accordance with Krylov et al. [29]. Motamedi et al. [30] assigned the two subpeaks to Al-Al (metallic Al) and Al-O (Al oxide) bonds appeared at 73.0 and 75.6 eV respectively. It can be noticed clearly that the peak position of the Al-N bond gets blue-shifted towards higher binding energy from 73.03 eV to73.74 eV with an increase in their FWHM values as the $N_2$ flow % is increased from 5 to 75%. But the integrated area under the peak keeps decreasing indicating less formation of Al-N bonds in the AlN films with an increase in the $N_2$ flow. Blue-shift is also observed for the Al-O bond due to charge transfer from Al to O but with the decrease in FWHM values from 3.43 to 1.74 eV. The increased area under the subpeak corresponding to the Al-O bond implies that the formation of the Al-O bond is enhanced with an increase in $N_2$ flow percentage. Table 5 shows the peak position and FWHM of deconvoluted Al-N and Al-O bond of Al(2p) core orbital of AlN thin films.

| $N_2$ flow rate (%) | Al-N | | Al-O | |
|---|---|---|---|---|
| | Peak position (in eV) | FWHM | Peak position (in eV) | FWHM |
| 5 | 73.03 | 1.56 | 74.35 | 3.43 |
| 17 | 73.03 | 1.55 | 74.25 | 2.43 |
| 25 | 73.13 | 1.60 | 74.45 | 2.17 |
| 30 | 73.23 | 1.62 | 74.55 | 2.08 |
| 40 | 73.44 | 1.72 | 74.76 | 1.94 |
| 75 | 73.74 | 1.63 | 74.96 | 1.74 |

[**Table 5** Peak position and FWHM of Al(2p) core orbital peaks for AlN films at various $N_2$ flow percentages.]

+

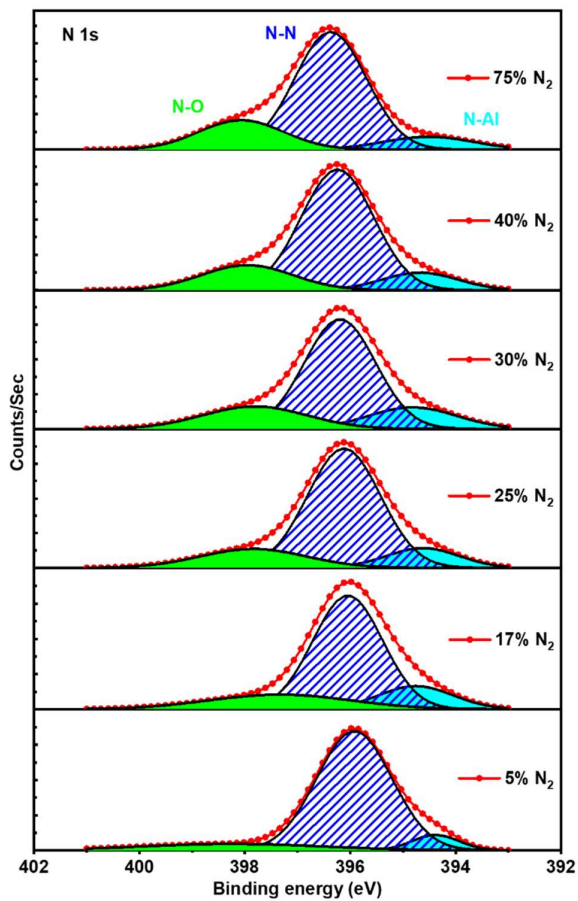

[**Fig. 5c.** Deconvolution of high resolution XPS spectra of N(1s) core orbital of AlN thin films at different N$_2$ flow rate.]

Fig. 5c shows the deconvolution of N(1s) core orbital spectra into three Gaussian-Lorentzian subpeaks of AlN films deposited at various N$_2$ flow percentages. The deconvoluted peaks that appeared at binding energies 394.62, 396.11 and 397.88 eV are assigned to N-Al (metallic), N-N (nitride) and N-O (oxide) bonds respectively and these bonds were located on the surface of the AlN films only. Due to sample handling N-Al bond has been shifted from 396 eV to 394 eV in the N (1s) core orbital spectra suggesting a modification in the electron density distribution around N atom bonded to Al. The area under the nitride peak covers more than 60% of the total area specifying N chemically attached to the N atom is abundant on the surface of AlN films compared

+

to the other bonds. An increase in the N-O bonding intensity was observed with increase in the N$_2$ flow rate for AlN films. The formation of an oxide layer on the surface of AlN films is confirmed to be increasing from the XPS results. The peak position and FWHM values of the N atom chemically attached to Al, N and O atoms are listed in Table 6 below.

| N2 flow rate (%) | N-Al | | N-N | | N-O | |
|---|---|---|---|---|---|---|
| | Peak position (in eV) | FWHM | Peak position (in eV) | FWHM | Peak position (in eV) | FWHM |
| 5 | 394.35 | 1.04 | 395.98 | 1.67 | 398.28 | 4.06 |
| 17 | 394.76 | 1.66 | 395.98 | 1.56 | 397.33 | 3.18 |
| 25 | 394.62 | 1.61 | 396.11 | 1.63 | 397.88 | 2.36 |
| 30 | 394.76 | 1.96 | 396.25 | 1.59 | 397.74 | 2.32 |
| 40 | 394.62 | 1.69 | 396.25 | 1.61 | 398.01 | 2.03 |
| 75 | 394.49 | 1.97 | 396.38 | 1.60 | 398.01 | 1.88 |

[Table 6 Peak position and FWHM of N(1s) core orbital peaks for AlN films at various N$_2$ flow percentages.]

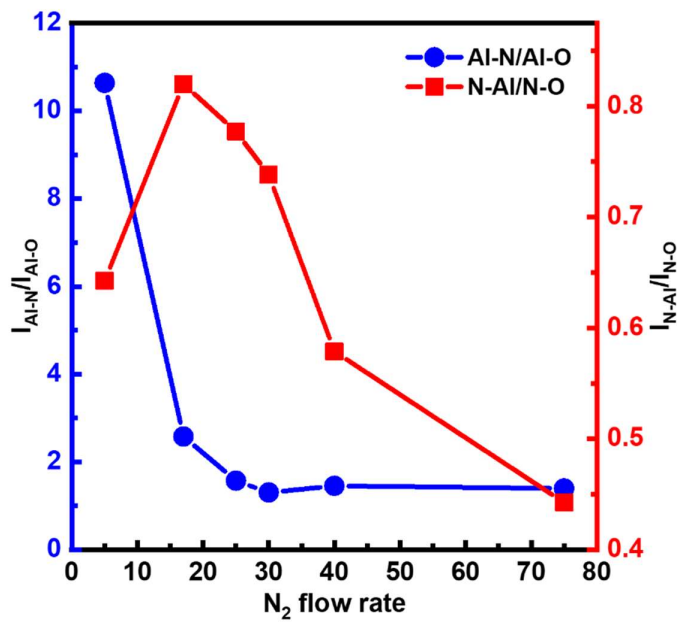

+

[**Fig. 5d.** Plot between the intensity of area ratios of Al-N to Al-O and N-Al to N-O vs. $N_2$ flow rate for AlN films.]

Fig. 5d shows the graph between the intensity of ratios between the area of Al nitride (Al-N) to Al oxide (Al-O) bonds of Al(2p) and N-Al (metallic) to N-O (oxide) bonds of N(1s) core orbital with respect to the $N_2$ flow rate for AlN thin films. The intensity of area ratios between Al-N and Al-O bonds keeps on decreasing exponentially up to 30% $N_2$ flow showing the exponential growth of Al-O bonds on the surface of AlN films which can be attributed to increased availability of oxygen species till 30% $N_2$ flow and then a slight increase is noticed which may be due to the saturation of available Al sites for further oxidation resulting in a slower rate of Al-O bond formation. And the intensity of area ratios between N-Al and N-O bonds of N(1s) core orbital increased for AlN films containing 5% to 17% $N_2$ flow leading to the increased incorporation of nitrogen into the film but however, the intensity then decreased drastically up to 75% $N_2$ flow. Beyond 17% $N_2$ flow, the excessive nitrogen incorporation started to have adverse effects. Though AlN films are nitrogen rich but at high $N_2$ flow rates, higher concentration of N-O bonds resulted relative to N-Al bonds in the AlN films.

**3.6 Raman analysis**

+

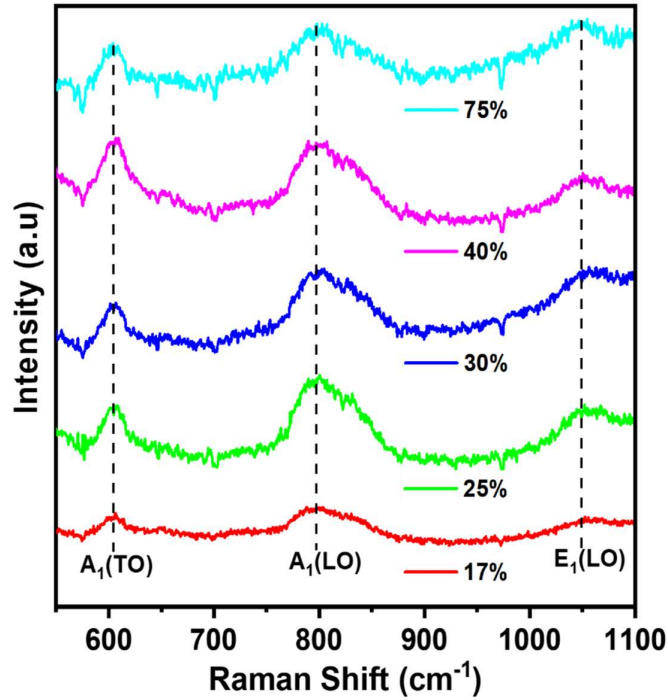

[**Fig. 6.** Raman spectra of AlN thin films with different $N_2$ flow.]

Raman spectroscopy is used to investigate the vibrational states of molecules and their local order. There remain only three acoustic and eight optic normal modes after attaining zero frequency modes for AlN [31]. The Raman spectra of AlN thin films with the variation in $N_2$ flow are shown in Fig. 6. $A_1$ and $E_1$ modes split into longitudinal optical (LO) and transverse optical (TO) components as they are optical modes. As AlN thin film is transparent, the laser can easily penetrate through the sample and the substrate signature is more likely to come out that is why quartz substrates were used to perform Raman spectroscopy. The Raman peaks of AlN in the energy range 600-900 cm$^{-1}$ are attributed to the optical phonons. The $A_1$ (TO), $A_1$ (LO) and $E_1$ (LO) modes range from 574.48–683.93 cm$^{-1}$, 752.44–875.13 cm$^{-1}$ and 926.40–1178.82 cm$^{-1}$ peaked at 604.29, 796.16 and 1049.23 cm$^{-1}$ respectively. The peak position of the $A_1$ (TO) mode comes out very close to the reported value of 611 cm$^{-1}$ [32]. For unstrained AlN,

+

the $A_1$ (LO), $E_1$ (TO) and $E_1$ (LO) modes usually appear at 890 cm$^{-1}$, 670.8 cm$^{-1}$ and 912 cm$^{-1}$ respectively [31]. The peak appearing at 800 cm$^{-1}$ is close to the reported value of $A_1$ (LO) mode and the peak at 1051 cm$^{-1}$ is assigned to $E_1$ (LO) mode. The blue shift of Raman peaks towards a higher wave number can be attributed to molecular interactions because of the presence of oxygen bonds in the AlN films which can lead to changes in the local environment affecting the vibrational energies. The increase in FWHM of the observed peaks demonstrates the deterioration of short-range order (SRO) in the thin films. The peak intensity for 25% N$_2$ flow is more indicating a higher phonon scattering effect at the grain boundaries. The phonon lifetime for individual phonons was calculated using the Heisenberg Uncertainty principle for energy and time as shown in table 7 below.

$$\Delta E . \Delta t \geq \frac{h}{4\pi} \tag{11}$$

The phonon lifetime for $A_1$(TO) and $A_1$(LO) modes are 20.58 ps and 9.13 ps respectively the longest for 25% N$_2$ flow and 13.13 ps for $E_1$(LO) mode which is the longest for the 40% N$_2$ flow. The intensity of the $A_1$(TO) and $A_1$(LO) modes gives information about the dielectric and mechanical properties respectively and their ratio is 1.06 which is the highest for 17% N$_2$ flow.

| RN$_2$(%) | Phonon lifetime (in ps) | | |
|---|---|---|---|
| | $A_1$(TO) | $A_1$(LO) | $E_1$(LO) |
| 17 | 14.75 | 8.69 | 4.37 |
| 25 | 20.58 | 9.13 | 12.67 |
| 30 | 15.52 | 7.01 | 10.78 |
| 40 | 17.62 | 7.12 | 13.13 |
| 75 | 12.43 | 5.48 | 10.72 |

[Table 7 Calculated phonon lifetime for the phonon modes of AlN thin films at different N$_2$ flow rates.]

## 4 Conclusion

+

The impact of nitrogen fraction on the structural properties, morphology, texture of the film, structural electronic environment and phonon vibration of the AlN thin films deposited by DC reactive magnetron sputtering has been addressed. XRD results proclaim that the c-axis orientation of AlN thin films increases with the increase in nitrogen flow percentage which is in accordance with the previously reported results. From XRR results it can be seen that for 40% $N_2$ flow the density of AlN thin films was the maximum which is 3.18 g/cm$^3$ and roughness was the minimum 2.766 A$^o$. Surface morphology analysis shows that the roughness average came out to be the highest for pure Al thin films and maximum RMS roughness for AlN thin films with 75% $N_2$ flow. Fractal factors have been calculated by fractal analysis of AFM images. XAS results are in good concurrence with the XRD confirming the c-axis-oriented growth. The shifting of feature-a from 407.58 eV to 406.96 eV with the increase in $N_2$ flow explains the presence of voids in AlN thin films. XPS results confirm the formation of an oxide layer on the surface of AlN films. The calculated film composition for 40% $N_2$ flow was 38.18 at.% Al and 61.81 at.% N. Short-range order deteriorates with the increase in FWHM of Raman peaks. The longest phonon lifetime was computed for $A_1$ (TO) mode which is 20.58 ps for AlN thin films at 25% $N_2$ flow.

+

+

**Acknowledgements**

The authors would like to thank Dr. Vasant Sathe and Mr. Ajay Rathore for carrying out the Raman spectroscopy experiments, Dr. R. Venkatesh and Mr. Mohan Kumar Gangrade for the AFM measurements at UGC-DAE CSR Indore, India and Mr. Rakesh Sah for SXAS measurements at Beam Line-01, Indus-2, RRCAT, Indore, India. Special thanks to Mr. Layanta Behera, Mr. Shailesh Kalal and Mr. Ashish Gupta for their invaluable support, guidance and discussions.

+

**Author contributions**

**Aishwarya Madhuri:** Investigation, Methodology, Formal analysis, Data curation, Writing - original draft, Writing - review & editing, Visualization. **Sanketa Jena:** Resources, Writing - review & editing, Visualization, Project administration. **Mukul Gupta:** Conceptualization**,** Validation, Resources, Writing - review & editing, Visualization, Funding acquisition, Supervision, Project administration. **Bibhu Prasad Swain:** Conceptualization, Validation, Resources, Writing - review & editing, Visualization, Supervision, Project administration.

**Funding**

The authors declare that there was no financial support from anywhere and anyone.

**Data availability**

Data sharing does not apply to this article as no datasets were generated or analyzed during the current study.

**Declarations**

**Conflict of Interest** The authors declare that they have no known competing financial interests or personal relationships that could have appeared to influence the work reported in this paper.